\documentclass[manyauthors]{fundam}
\usepackage{tikz-cd}
\usepackage{yfonts}
\usepackage{amsmath,amssymb,amsfonts}
\usepackage{float}
\usepackage{enumerate}
\usepackage{tikz}
\usepackage{dirtytalk}
\newtheorem{thm}{Theorem}
\newtheorem{lem}{Lemma}
\newtheorem{prop}{Proposition}
\newtheorem{cor}{Corollary}
\newtheorem{defn}{Definition}

\newtheorem{rem}{Remark}
\newtheorem{note}{Note}

\begin{document}
	
\title{Coalgebraic Fuzzy geometric logic	}

\author{Litan Kumar Das\thanks{ld06iitkgp@gmail.com}\\
	Department of Mathematics \\
	Jadavpur University, Kolkata, India
	\and Kumar Sankar Ray\thanks{ksray@isical.ac.in}\\
	ECSU, Indian Statistical Institute, Kolkata, India \\
\and Prakash Chandra Mali\thanks{pcmali1959@gmail.com}\\
Department of Mathematics \\
Jadavpur University, Kolkata, India
}

	
	






\maketitle
\address{Litan Kumar Das, Department of Mathematics, Jadavpur University}
\runninghead{Das, L K., Ray, K S., \& Mali, P C}{Coalgebraic Fuzzy geometric logic}
\begin{abstract}
The paper aims to develop a framework for coalgebraic fuzzy geometric logic by adding modalities to the language of fuzzy geometric logic. Using the methods of coalgebra, the modal operators are introduced in the language of fuzzy geometric logic. To define the modal operators, we introduce a notion of fuzzy-open predicate lifting. Based on coalgebras for an endofunctor $T$ on the category $\textbf{Fuzzy-Top}$ of fuzzy topological spaces and fuzzy continuous maps, we build models for the coalgebraic fuzzy geometric logic. Bisimulations for the defined models are discussed in this work.
\end{abstract} 

\section{Introduction}
Fuzzy geometric logic has been introduced in \cite{chakraborty2017fuzzy} which is a natural generalization of the propositional geometric logic \cite{vickers1996topology}. 
Vicker in \cite{vickers1996topology} developed geometric logic based on point-free topology, logic, and the logic of finite observations \cite{abramsky2011domain}. It has been pointed out in several works (see\cite{goldblatt2014topoi,johnstone2014topos,maclane2012sheaves,vickers1993geometric,vickers1999topology}). The language of geometric logic is formed on a set of propositional variables by applying propositional connectives: finite conjunction$(\wedge)$ and arbitrary disjunction$(\bigvee)$. The property of finite observability is preserved by these connectives. Vickers in \cite{vickers1996topology} explored the relationship between topological spaces, topological systems and geometric logic. A topological system is defined by a triple $(X,\models, A)$ in which $X$ occurs as a non-empty set of objects, $A$ defines a  frame and $\models$ is given by a satisfaction relation from $X$ to $A$. 
The authors in \cite{chakraborty2017fuzzy} have generalized geometric logic to the many-valued context by extending the notion of satisfiability relation. They observed that if the satisfaction relation is fuzzy then the corresponding consequence relation has two possibilities: crisp or fuzzy. Consequently, they introduced general fuzzy geometric logic and fuzzy geometric logic with graded consequence. The relationship among fuzzy geometric logic, fuzzy topology and fuzzy topological systems has also been shown in their work.
The concept of Fuzzy topological spaces introduced in \cite{chang1968fuzzy} and it was taken into account in various works(e.g.,\cite{chakraborty1992fuzzy,hohle1986fuzzy,lowen1976fuzzy,pao1980fuzzy,syropoulos2011fuzzy}).
Graded consequence and its related aspects can be found in \cite{chakraborty2019theory}.
Different kinds of modal logics were developed based on the theory of coalgebras. In \cite{kurz2006coalgebras}, A. Kurz discussed coalgebraic logic in more detail. Coalgebraic logic for functors on the category of sets extensively studied in \cite{cirstea2011modal,kupke2011coalgebraic}, where they used the notion of predicate lifting \cite{pattinson2003coalgebraic} or relation lifting \cite{moss1999coalgebraic} to define modal operators. In \cite{enqvist2014generalized}, Enqvist and Sourabh developed coalgebraic modal logic in the category of Stone coalgebras.
Coalgebraic geometric logic has been developed in \cite{venema2013generalised,bezhanishvili2019coalgebraic}, where modal operators are defined to the language of geometric logic based on the theory of coalgebraic logic. Motivated by their work, we study modalities in fuzzy geometric logic using the coalgebraic machinery of predicate liftings. The structures, named fuzzy geometric models for $T$, provide the semantics for coalgebraic fuzzy geometric logic.
The notion of bisimulation \cite{sangiorgi2009origins} is extensively applied in various fields of computer science as well as in mathematics. The bisimulation between coalgebras is a fundamental concept in ``state-based systems'' which associates the states in systems with the same behaviour. In \cite{staton2011relating}, Staton developed different notions of coalgebraic bisimulation and studied the relationship between these notions. Bisimulations for the Stone coalgebras studied in \cite{enqvist2014generalized}. In our work, we have discussed bisimulations for the proposed fuzzy geometric models. It is expected that the proposed coalgebraic logic and the results developed in this work will have implications in computer science areas such as knowledge representations, logic programming, formal verification, and fuzzy reasoning.
The paper is organized as follows.
In Section \ref{p0}, we have mentioned the ideas that are required for the work. We define coalgebraic logic for \textbf{Fuzzy-Top}-coalgebras in Section \ref{c0}, and then introduce fuzzy geometric models for an endofunctor $T$. In Section \ref{fm}, we construct a final model in the category $FMOD(T)$ of fuzzy geometric models for $T$, where $T$ is an endofunctor on the category \textbf{SFuzzy-Top} of sober fuzzy topological spaces. In Section \ref{bf}, we discuss bisimulations for fuzzy geometric models. We conclude the paper by mentioning future directions of work in Section \ref{f}.

\section{Preliminaries}
\label{p0}
The reader is referred to \cite{adamek1990h} for category theory basics. For clarity and understanding of our work, we mention some useful ideas.\\
In 1965, Zadeh investigated fuzzy set theory \cite{ZADEH1965338}. We go through some fundamental ideas in fuzzy set theory.
\begin{defn}
	A fuzzy set $\tilde{f}$ on a set $S$ is defined by the membership function $\tilde{f}:S\to [0,1]$.
\end{defn}
Let $\tilde{f}^c$ denote the complement of $\tilde{f}$. Define $\tilde{f}^c: S\to [0,1]$ by $\tilde{f}^c(s)=1-\tilde{f}(s)$, $\forall s\in S$. $\tilde{f}^c$ is a fuzzy set on $S$.
\begin{note}
	If $\tilde{f_1}$ and $\tilde{f_2}$ are fuzzy sets on $S$, then $\tilde{f_1}\vee\tilde{f_2}$ and $\tilde{f_1}\wedge\tilde{f_2}$ are fuzzy sets on $S$, where the fuzzy sets $\tilde{f_1}\vee\tilde{f_2}$ and $\tilde{f_1}\wedge\tilde{f_2}$ are defined by $(\tilde{f_1}\vee\tilde{f_2})(s)=\tilde{f_1}(s)\vee\tilde{f_2}(s)$ and $(\tilde{f_1}\wedge\tilde{f_2})(s)=\tilde{f_1}(s)\wedge\tilde{f_2}(s)$, respectively.
\end{note}
\begin{rem}
For each $s\in S$, the grade of membership of $s$ in the fuzzy set $\tilde{f}$ is given by the value $\tilde{f}(s)$. It is represented by the symbol $gr(s\in\tilde{f})$.
\end{rem}
\begin{defn}
	Let $S_1$ and $S_2$ be two sets and $f:S_1\to S_2$ be a given function. For a fuzzy set $\tilde{s_1}$ on $S_1$, the direct image $f(\tilde{s_1}):S_2\to [0,1]$ of the fuzzy set $\tilde{s_1}$ under the function $f$ is defined by $f(\tilde{s_1})(s)=\bigvee \{\tilde{s_1}(t):t\in f^{-1}(\{s\})\}$, where $s\in S_2$.
\end{defn}
\begin{defn}
	Let $S_1$ and $S_2$ be two sets and $f:S_1\to S_2$ be a given function. For a fuzzy set $\tilde{s_2}$ on $S_2$, the inverse image $f^{-1}(\tilde{s_2}): S_1\to [0,1]$ of the fuzzy set $\tilde{s_2}$ under the function $f$ is defined by $f^{-1}(\tilde{s_2})=\tilde{s_2}\circ f$. 
	\end{defn}
\begin{defn}
	Let $\mu$ and $\eta$ be fuzzy sets on $S$. Then, $\mu$ is a fuzzy subset of $\eta$, denoted by $\mu\leq \eta$, $\iff$ $\mu(s)\leq \eta(s)$, $\forall s\in S$.
\end{defn}
\begin{defn}
	\label{IVP}
	Consider a mapping $f:S\to T$ and a collection $\{\mu_i:i\in \mathcal{I}, \mathcal{I} \text{ is an index set }\}$ of fuzzy sets on $T$. Then 
	\begin{enumerate}[(i)]
		\item $f^{-1}(\displaystyle\bigvee_{i\in \mathcal{I}} \mu_i)=\bigvee_{i\in \mathcal{I}}f^{-1}(\mu_i)$;
		\item $f^{-1}(\displaystyle\bigwedge_{i\in \mathcal{I}} \mu_i)=\bigwedge_{i\in \mathcal{I}}f^{-1}(\mu_i)$.
	\end{enumerate}
\end{defn}
We recall the definition of fuzzy topological spaces from \cite{chang1968fuzzy}.
\begin{defn}
	Let $S$ be a set. A collection $\tau_S$ of fuzzy sets on $S$ is said to be fuzzy topology on $S$ if the following conditions hold:
	\begin{enumerate}[(i)]
		\item $\tilde{\emptyset}, \tilde{S}\in \tau_S$, where $\tilde{\emptyset}(s)=0$, $\forall s\in S$ and $\tilde{S}(s)=1$, $\forall s\in S$;
		\item if $\tilde{g_1},\tilde{g_2}\in \tau_s$ then $\tilde{g_1}\wedge\tilde{g_2}\in\tau_S$, where $(\tilde{g_1}\wedge\tilde{g_2})(s)=\tilde{g_1}(s)\wedge\tilde{g_2}(s)$;
		\item if $\tilde{g_{j}}\in \tau_S$ for $j\in \Lambda$, $\Lambda$ is an index set, then $\displaystyle\bigvee_{j\in\Lambda} \tilde{g_{j}}\in\tau_S$, where $\displaystyle\bigvee_{j\in\Lambda} \tilde{g_{j}}(s)=\sup_{j\in \Lambda}\{\tilde{g_{j}}(s)\}$.
	\end{enumerate}
\end{defn}
Then, the pair $(S,\tau_S)$ is referred to as a fuzzy topological space and members of $\tau_S$ are said to be fuzzy open sets on $(S,\tau_S)$. 
\begin{defn}[\cite{goguen1973fuzzy}]
	Let $(S,\tau_S)$ be a fuzzy topological space. Then a subset $\mathfrak{B}$ of $\tau_S$ is called a basis for $(S,\tau_S)$ if it satisfies the following conditions:
	\begin{enumerate}[(i)]
		\item if $\tilde{b}_1,\tilde{b}_2\in\mathfrak{B}$ then $\tilde{b}_1\wedge\tilde{b}_2\in\mathfrak{B}$;
		\item for each member $\tilde{t}\in\tau_S$, there exists a subcollection $\mathcal{C}=\{\tilde{t}_{j}\in\mathfrak{B}:j\in\Lambda\}$ such that $\tilde{t}=\displaystyle\bigvee_{j\in\Lambda}\tilde{t}_{j}$.
	\end{enumerate}
\end{defn}
\begin{defn}[\cite{goguen1973fuzzy}]
	Let $\tau_S$ be a fuzzy topology on $S$ and $\mathfrak{S}\subset \tau_S$. Then, $\mathfrak{S}$ is a subbasis for a  fuzzy space $(S,\tau_S)$ $\iff$ the collection of all finite meets of members of $S$ is a basis for $(S,\tau_S)$.
\end{defn}
\begin{defn}
	A fuzzy topological space $(S,\tau_S)$ is said to be Kolmogorov space or $T_0$-space if for any pair $(x,y)$ of distinct points in $S$, there is a fuzzy open set $\tilde{g}$ on $S$ such that $\tilde{g}(x)\neq\tilde{g}(y)$.
\end{defn}
\begin{defn}
Let $(F,\tau_{F})$ and $(G,\tau_{G})$ be fuzzy topological spaces. A mapping $f:F\to G$ is fuzzy continuous if and only if, for every fuzzy open set $\tilde{g}$ on $(G,\tau_G)$, $f^{-1}(\tilde{g})$ is a fuzzy open set on $(F,\tau_F)$.
\end{defn}
Definitions of category, opposite category and functors can be found in \cite{adamek1990h}. We recall the concept of natural transformations.
\begin{defn}[\cite{adamek1990h}]
	Let $\mathcal{F}$ and $\mathcal{G}$ be functors from a category $\mathcal{A}$ to a category $\mathcal{B}$. A natural transformation $\mathcal{T}:\mathcal{F}\to\mathcal{G}$ is a mapping that carries each object $A$ in $\mathcal{A}$ to a morphism $\mathcal{T}_A:\mathcal{F}(A)\to \mathcal{G}(A)$ in $\mathcal{B}$ such that for any morphism $\zeta:A\to \bar{A}$ in $\mathcal{A}$ we have $\mathcal{G}(\zeta)\circ \mathcal{T}_A=\mathcal{T}_{\bar{A}}\circ \mathcal{F}(\zeta)$. As a result, the Fig. \ref{NAT} commutes.
	\begin{figure}[H]
			\begin{center}
	\begin{tikzcd}
\mathcal{F}(A) \arrow[rr, "\mathcal{T}_A"] \arrow[d, "\mathcal{F}(\zeta)"'] &  & \mathcal{G}(A) \arrow[d, "\mathcal{G}(\zeta)"] \\
\mathcal{F}(\bar{A}) \arrow[rr, "\mathcal{T}_{\bar{A}}"']                   &  & \mathcal{G}(\bar{A})\end{tikzcd}                     
\end{center}
	\caption{Illustration of natural transformation}
	\label{NAT}
\end{figure}
\end{defn}
Goguen first considered the category of fuzzy sets in \cite{goguen1967fuzzy}. Several authors studied on the category of fuzzy sets (e.g.,\cite{walker2004categories,barr1986fuzzy,goguen1969categories}). Let $\textbf{FS}$ denote the category of fuzzy sets.
\begin{defn}[\cite{walker2004categories}]
The category $\textbf{FS}$ is defined as follows:
\begin{enumerate}[(i)]
	\item An object in $\textbf{FS}$ is a pair $(S,\tilde{g})$, where $S$ is a set and $\tilde{g}:S\to [0,1]$ is a membership function;
	\item A morphism $f:(S,\tilde{g})\to (T,\tilde{h})$ in $\textbf{FS}$ is a function $f:S\to T$ such that $\tilde{g}(s)\leq f^{-1}(\tilde{h})(s)$.	
\end{enumerate}
\end{defn}
Let $\textbf{Fuzzy-Top}$ denote the category of fuzzy topological spaces.
\begin{defn}
	The category $\textbf{Fuzzy-Top}$ is defined as follows:
	\begin{enumerate}[(i)]
		\item Objects in $\textbf{Fuzzy-Top}$ are fuzzy topological spaces $(S,\tau_S)$;
		\item Morphisms $f:(S,\tau_S)\to (T,\tau_T)$ in $\textbf{Fuzzy-Top}$ are fuzzy continuous mappings. 
	\end{enumerate}
\end{defn}

\begin{defn}
\label{p1} 
A functor $\mathcal{Q}$ from the category $\textbf{Fuzzy-Top}$ to the category $\textbf{FS}$ of fuzzy sets can be defined as follows:
\begin{enumerate}[(i)]
\item For an object $(S,\tau_S)$ in \textbf{Fuzzy-Top}, define $\mathcal{Q}(S)=\text{set of fuzzy open sets on } (S,\tau_S)$;
\item For a  morphism $\phi:(S,\tau_S)\to (T,\tau_T)$ in \textbf{Fuzzy-Top}, define $\mathcal{Q}(\phi)=\phi^{-1}:\mathcal{Q}(T)\to \mathcal{Q}(S)$ by $\phi^{-1}(\mu)=\mu\circ\phi$, $\mu\in \mathcal{Q}(T)$.
\end{enumerate}
\end{defn}
\begin{defn}[\cite{vickers1996topology}]
A frame is a partially ordered set which is closed under finite meet $(\wedge)$ and arbitrary join $(\bigvee)$ such that $\wedge$ distributes over $\bigvee$.
\end{defn}
\begin{note}
	Let $(S,\tau_S)$ be a fuzzy topological space. Then, the fuzzy topology $\tau_S$ on $S$ can be considered as a frame.
	
\end{note}
\begin{defn}[\cite{vickers1996topology}]\label{frp}
	A function $f$ from a frame $F_1$ to a frame $F_2$ is said to be a frame homomorphism if the function $f$ preserves finite meets and arbitrary joins.
\end{defn}
The collection of frames and frame homomorphisms forms a category, denoted by $\textbf{FRM}$. \\
Similar to other algebraic structures, frames may be presented by generators and relations $\langle G \vline R \rangle$, where $G$ denotes the set of generators, and $R$ is the set of relations between expressions generated by $G$. One can find a detailed description of frame presentations in \cite{vickers1999topology}.
\begin{note}[\cite{bezhanishvili2019coalgebraic}]\label{fpr}
	Consider a frame $F_1$. Now, $\langle G\vline R \rangle$ presents the frame $F_1$ if $\exists$ an assignment $h: G \to F_1^*$, where $F_1^*$ denotes the underlying set of $F_1$, such that the following properties hold:
	\begin{enumerate}[(i)]
		\item $F_1$ is generated by the set $\{h(s):s\in G\}$;
\item[]$h$ can be extended to an assignment $\hat{h}$ for any expression $r$ that is generated by $G$.
	\item If $r*=r'*$ is a relation in $R$, then $\hat{h}(r*)=\hat{h}(r'*)$ in $F_1$;
	\item For a frame $F_2$ and an assignment $h': G\to F_2^*$ that satisfies $(ii)$ there is a unique frame homomorphism $g:F_1\to F_2$ such that $g^*\circ h=h'$, where $g^*$ is a mapping from $F_1^*$ to $F_2^*$. So, the diagram shown in Fig. \ref{FPR} commutes.
			\begin{figure}[H]
				\begin{center}
		\begin{tikzcd}
			G \arrow[d, "h'"'] \arrow[r, "h"] & F_1^* \arrow[ld, "g^*"] \\
			F_2^*                             &                        
		\end{tikzcd}
	\end{center}
	\caption{Illustration of frame presentation}
	\label{FPR}
	\end{figure}
\end{enumerate}
\end{note}
\begin{rem}\label{spr}
We can define a frame homomorphism $f:F_1\to F_2$ from a frame $F_1$ to a frame $F_2$, where $\langle G\vline R \rangle$ presents the frame $F_1$. According to Note \ref{fpr}, it is sufficient to provide an assignment $\hat{f}:G\to F_2$ that satisfies the condition that if $r*=r'*$ is a relation in $R$, then $\hat{f}(r*)=\hat{f}(r'*)$ in $F_2$.
\end{rem}
Coalgebras are categorical structures that are dual or opposite (in the sense of category theory) to the notion of algebras. The coalgebraic approach is abundantly applied in computer science and artificial intelligence (e.g., knowledge representation, concurrency, logical reasoning, automata theory, etc.). 
\begin{defn}[\cite{adamek2005introduction}]
Assume that $T$ is an endofunctor on a category $\mathcal{S}$. A $T$-coalgebra is a pair $(A,\zeta)$, where $A$ is an object in $\mathcal{S}$ and $\zeta:A\to T(A)$ is a morphism in $\mathcal{S}$.
\end{defn}
\begin{defn}[\cite{adamek2005introduction}]
A morphism between $T$-coalgebras $(A,\delta)$ and $(B,\beta)$ is defined by a morphism $\psi:A\to B$ in $\mathcal{S}$ satisfying the equation $T(\psi)\circ \delta=\beta\circ\psi$, i.e., the diagram shown in Fig. \ref{cm} commutes.
\begin{figure}[H]
	\begin{center}
\begin{tikzcd}
	A \arrow[r, "\psi"] \arrow[d, "\delta"'] & B \arrow[d, "\beta"] \\
	T(A)  \arrow[r, "T(\psi)"']              & T(B)                
\end{tikzcd}
\end{center}
\caption{Illustration of coalgebra morphism}
\label{cm}
\end{figure}
\end{defn}
$T$-coalgebras and morphisms between $T$-coalgebras form a category, denoted by $\textbf{COALG}(T)$.\\
A final coalgebra is a final or terminal object in $\textbf{COALG}(T)$. It has a significant impact on computer science. The final coalgebra is crucial as it makes sense of behaviourally equivalent states in coalgebras. 
\begin{defn}
A final coalgebra in $\textbf{COALG}(T)$ is a $T$-coalgebra $(A,\delta)$ which satisfies that for each $T$-coalgebra $(B,\beta)$, a unique morphism exists from $(B,\beta)$ to $(A,\delta)$.
\end{defn}
\begin{defn}
Let $(A,\delta)$ and $(B,\beta)$ be objects in $\textbf{COALG}(T)$. We say that any two states $a\in A$ and $b\in B$ are behaviourally equivalent if there exists an object $(C,\alpha)$ in $\textbf{COALG}(T)$
and $T$-coalgebra morphisms $g:(A,\delta)\to (C,\alpha)$ and $h:(B,\beta)\to (C,\alpha)$ such that $g(a)=h(b)$.
\end{defn}

\begin{defn}[\cite{aczel1989final}]
 Let $(A,\delta)$ and $(B,\beta)$ be two $T$-coalgebras. Then a relation $\mathcal{R}\subseteq A\times B$ is said to be a bisimulation between $(A,\delta)$ and $(B,\beta)$ if there exists a $T$-coalgebra $(\mathcal{R},\gamma)$ such that the projection maps $\pi_1:\mathcal{R}\to A$ and $\pi_2:\mathcal{R}\to B$ are coalgebra morphisms and satisfy the relations   $\delta\circ\pi_1=T(\pi_1)\circ\gamma$, $\beta\circ\pi_2=T(\pi_2)\circ\gamma$. So, the diagram shown in Fig. \ref{bsm} is commutative.
	\begin{figure}[H]
		\begin{center}
		\begin{tikzcd}
				A \arrow[d, "\delta"'] &  & \mathcal{R} \arrow[ll, "\pi_1"'] \arrow[d, "\gamma"] \arrow[rr, "\pi_2"] &  & B \arrow[d, "\beta"] \\
				T(A)                   &  & T(\mathcal{R}) \arrow[ll, "T(\pi_1)"] \arrow[rr, "T(\pi_2)"']            &  & T(B)                
			\end{tikzcd}
			\end{center}
			\caption{Illustration of coalgebraic bisimulation}
			\label{bsm}
	\end{figure}
	\end{defn}

\begin{defn}
We define a functor $\mathcal{P}:\textbf{Fuzzy-Top}\to \textbf{FRM}$ as follows:
\begin{enumerate}[(i)]
	\item For an object $S$ in \textbf{Fuzzy-top}, define $\mathcal{P}(S)=\tau_S$;
	\item For an arrow $\eta: S_1\to S_2$ in \textbf{Fuzzy-top}, define $\mathcal{P}(\eta):\mathcal{P}(S_2)\to\mathcal{P}(S_1)$ by $\mathcal{P}(\eta)(\xi)=\xi\circ \eta$, where $\xi\in\mathcal{P}(S_2)$.
\end{enumerate}
\end{defn}
\begin{note}
	Let $F$ be a frame and $(S,\tau_S)$ be a fuzzy topological space. Then,
	the frame $F$ is spatial if there is an isomorphism from $F$ to $\mathcal{P}(S)$.
\end{note}
Consider that $\textbf{S-FRM}$ is the category of spatial frames and homomorphisms between frames. Let $F$ denote a frame and $\textit{PT}F$ denote the collection of frame homomorphisms $h$ from $F$ to $[0,1]$. Then the collection $\{\Psi(a): a\in F\}$ is a fuzzy topology on $\textit{PT}F$, where for each $a\in F$,  $\Psi(a)$ is a membership function from $\textit{PT}F$ to $[0,1]$ which is defined by $\Psi(a)(h)=h(a)$. 
\begin{defn}[\cite{kotze1997fuzzy}]
Let $\tau$ be a fuzzy topology on $S$. Assume that $\tilde{T}\in\tau$. A membership function $\Psi(\tilde{T}):\textit{PT}\tau\to [0,1]$ can be defined as $\Psi(\tilde{T})(h)=h(\tilde{T})$. Then, $\Psi(\tilde{T})$ is a fuzzy set on $\textit{PT}\tau$.
\end{defn}
The collection $\{\Psi(\tilde{T}):\tilde{T}\in\tau\}$ is a fuzzy topology on $\textit{PT}\tau$.
\begin{cor}[\cite{kotze1997fuzzy}]
Consider a fuzzy topological space $(S,\tau)$. A mapping $f:S\to \textit{PT}\tau$ is defined by $f(s)(\tilde{\phi})=\tilde{\phi}(s)$, where $\tilde{\phi}\in \tau$. Hence, $(S,\tau)$ becomes sober $\iff$ $f$ is bijective.
\end{cor}
The category of sober fuzzy topological spaces and fuzzy continuous maps is denoted by \textbf{SFuzzy-Top}.
\begin{defn}
	We define a functor $\textit{PT}:\textbf{FRM}\to \textbf{Fuzzy-Top}$ as follows:
	\begin{enumerate}[(i)]
		\item For an object $F$ in \textbf{FRM}, define $\textit{PT}(F)=\textit{PT}F$;
		\item For an arrow $f:F\to F'$ in $\textbf{FRM}$, define $\textit{PT}(f):\textit{PT}F'\to\textit{PT}F$ by $\textit{PT}(f)(h)=h\circ f$, where $h\in\textit{PT}F'$.
	\end{enumerate}
	
\end{defn}

\begin{thm}
	\label{s0}
	The category $\textbf{S-FRM}$ is dually equivalent to the category \textbf{SFuzzy-Top}.
	\begin{proof}
		Let $id_1$ and $id_2$ be the identity functors on $\textbf{S-FRM}$ and \textbf{SFuzzy-Top}, respectively.
		We define two natural transformations $\zeta:id_1\to \mathcal{P}\circ\textit{PT}$ and $\eta:id_2\to\textit{PT}\circ\mathcal{P}$. For a spatial frame $F$, we define $\zeta_{F}:F\to \mathcal{P}\circ\textit{PT}(F)$ by $\zeta_{F}(u)(h)=h(u)$, where $h\in \textit{PT}F$. Since $F$ is a spatial frame, we have $\zeta_{F}$ is an isomorphism. It becomes easy to observe that $\zeta$ is a natural transformation. As a result, $\zeta$ is a natural isomorphism.\\
		For an object $S$ in \textbf{SFuzzy-Top}, define $\eta_S:S\to\textit{PT}\circ\mathcal{P}S$ by $\eta_S(s)(\tilde{g})=\tilde{g}(s)$, $\forall s\in S$ and $\tilde{g}\in\mathcal{P}(S)$. As $S$ is sober, so $\eta_S$ is bijective. We observe that, for $\tilde{g}\in\mathcal{P}(S)$, $\eta_S^{-1}(\Psi(\tilde{g}))(s)=\Psi(\tilde{g})(\eta_S(s))=\eta_S(s)(\tilde{g})=\tilde{g}(s)$. Therefore, $\eta_S^{-1}(\Psi(\tilde{g}))=\tilde{g}$. Moreover, $\eta_S$ is an open map because $\eta_S(\tilde{g})(h)=\bigvee\{\tilde{g}(s):s\in\eta_S^{-1}(h)\}=h(\tilde{g})=\Psi(\tilde{g})(h)$. Therefore, $\eta_S(\tilde{g})=\Psi(\tilde{g})$. Consequently, $\eta_S$ is a fuzzy homeomorphism. It can be shown that $\eta$ is a natural transformation. Hence, $\eta$ is a natural isomorphism.
	\end{proof}
\end{thm}

\section{Coalgebraic logic }
\label{c0}
             For this section, $T$ is assumed to be an arbitrary endofunctor on the category $\mathcal{C}=\textbf{Fuzzy-Top}$. We define coalgebraic logic for \textbf{Fuzzy-Top}-coalgebras. First, we introduce a notion of a predicate lifting for the endofunctor $T$, called fuzzy-open predicate lifting. 
\begin{defn}
A fuzyy-open predicate lifting is defined by a natural transformation 
\begin{equation*}
\lambda:\mathcal{Q}^n\to \mathcal{Q}\circ T.
\end{equation*}
The dual of $\lambda$ is denoted by $\breve{\lambda}$ and defined as $\breve{\lambda}(\mu_1,\mu_2,\cdots, \mu_n)=1-\lambda(1-\mu_1, 1-\mu_2,\cdots, 1-\mu_n)$, where $\mu_i\in\mathcal{Q}(S)$, $i=1,2,\cdots,n$ and $S$ is an object in $\textbf{Fuzzy-Top}$.
\end{defn}
\begin{defn}
\label{p2}
We call the fuzzy-open predicate lifting $\lambda$ is
\begin{enumerate}[(i)]
\item monotone if for every object $S$ in \textbf{Fuzzy-Top} and $\mu_i,\eta_i\in\mathcal{Q}(S)$, $i=1,2,\cdots, n$ such that $\mu_1\leq \eta_1,\cdots, \mu_n\leq\eta_n \Rightarrow \lambda_{S}(\mu_1,\cdots,\mu_n)\leq\lambda_{S}(\eta_1,\cdots,\eta_n)$.
\end{enumerate}
\end{defn}
Let $\Sigma$ be a collection of fuzzy-open predicate liftings for $T$. Then, the collection $\Sigma$ is said to be a fuzzy geometric modal signature for $T$. If every member of $\Sigma$ is monotone, then the fuzzy geometric modal signature for $T$ is said to be monotone.
\begin{defn}\label{p4}
 A fuzzy geometric modal signature for an endofunctor $T:\textbf{Fuzzy-Top}\to \textbf{Fuzzy-Top}$ is said to be characteristic for $T$ if for each object $S$ in $\textbf{Fuzzy-Top}$, the collection $\{\lambda_S(\mu_1,\cdots,\mu_n): \lambda\in\Sigma, \mu_i\in\mathcal{Q}(S)\}$ is a sub-basis for the fuzzy topology on $TS$.
\end{defn}
We now introduce the modal language for the fuzzy geometric modal signature $\Sigma$.
\begin{defn}
\label{p5}
For a given fuzzy geometric modal signature $\Sigma$ for $T:\textbf{Fuzzy-Top}\to \textbf{Fuzzy-Top}$, we define the modal language $\textbf{FGML}(\Sigma)$ by the grammar\\
    $\beta :: =\top\lvert p \rvert \beta_1\wedge\beta_2 \lvert \displaystyle\bigvee_{j\in J}\beta_j \rvert \heartsuit^{\lambda}(\beta_1,\beta_2,\cdots,\beta_n)$, with $\lambda\in\Sigma$, $\Phi$ denotes the set of propositional variables $p$ and $J$ is an index set. 
\end{defn}
\begin{defn}
\label{p6}
A fuzzy geometric model for the functor $T$ is defined by $\mathcal{S}=(S,\sigma,\mathcal{V})$ in which $(S,\sigma)$ is a $T$-coalgebra, and $\mathcal{V}:\Phi\to\mathcal{Q}(S)\subseteq [0,1]^S$ is a valuation function.
\end{defn}
\begin{defn}\label{p8}
A category $FMOD(T)$ is defined as follows:
\begin{enumerate}
\item Every object in $FMOD(T)$ is a fuzzy geometric model for $T$;
\item A morphism $f:(S,\sigma_1,\mathcal{V}_S)\to (S',\sigma_2,\mathcal{V}'_{S'})$ in $FMOD(T)$  is a coalgebra morphism $f:(S,\sigma_1)\to (S',\sigma_2)$ which satisfies the condition:
 $f^{-1}\circ \mathcal{V}'_{S'}=\mathcal{V}_S$.
\end{enumerate}
\end{defn}
\begin{defn}
\label{p9}
The semantics of a formula $\alpha$ in \textbf{FGML$(\Sigma)$} on a fuzzy geometric model $\mathcal{S}=(S,\sigma,\mathcal{V})$ is defined as:
\begin{enumerate}[(i)]
\item $[[\top]]_{\mathcal{S}}(s)=1$;
\item $[[p]]_{\mathcal{S}}(s)=\mathcal{V}(p)(s)$;
\item $[[\alpha\wedge \beta]]_{\mathcal{S}}(s)=[[\alpha]]_{\mathcal{S}}(s)\wedge [[\beta]]_{\mathcal{S}}(s)$;
\item $[[\displaystyle\bigvee_{i\in J}\alpha_i]]_{\mathcal{S}}(s)=Sup\{[[\alpha_i]]_{\mathcal{S}}(s)\}$;
\item $[[\heartsuit^{\lambda}(\alpha_1,\alpha_2,\cdots,\alpha_n)]]_{\mathcal{S}}(s)=\lambda_S([[\alpha_1]]_{\mathcal{S}},[[\alpha_2]]_{\mathcal{S}},\cdots,[[\alpha_n]]_{\mathcal{S}})\circ\sigma(s)$.
\end{enumerate}
\end{defn}
Grade of a formula $\alpha$ in \textbf{FGML$(\Sigma)$} satisfied by a state $s$ in $S$ is denoted by $gr(s\models\alpha)$ and defined by $gr(s\models\alpha)=[[\alpha]]_{\mathcal{S}}(s)$. Two states $s$ and $t$ in $S$ are modally equivalent if $gr(s\models \alpha)=gr(t\models \alpha)$, for all formulas $\alpha$ in $\textbf{FGML}(\Sigma)$. We express it by the notation $s\equiv_{\Sigma} t$. 
\begin{defn}
	\label{BV}
	Let $\mathcal{B}=(B,\sigma_1,\mathcal{V}_B)$ and $\mathcal{B}'=(B',\sigma_2,\mathcal{V}_{B'})$ be fuzzy geometric models for $T$. States $b\in B$ and $b'\in B'$ are said to be behaviourally equivalent in $FMOD(T)$ if there exists an object $\mathcal{C}=(C,\gamma,\mathcal{V}_C)$ in $FMOD(T)$ and morphisms $g:\mathcal{B}\to\mathcal{C}$ and $h:\mathcal{B}'\to \mathcal{C}$ in $FMOD(T)$ such that $g(b)=h(b')$.
\end{defn}
In the following Proposition \ref{p10}, we shall show that fuzzy geometric model morphisms preserve truth degrees.

\begin{prop}
\label{p10}
Assume that $f:\mathcal{S}=(S,\sigma_1,\mathcal{V}_S)\to \mathcal{K}=(K,\sigma_2,\mathcal{V}_K)$ is a morphism between fuzzy geometric models. Then $gr(s\models \alpha)=gr(f(s)\models \alpha)$, $\forall \alpha\in \textbf{FGML}(\Sigma)$ and $s\in S$.
\begin{proof}
We have to show that $[[\alpha]]_{\mathcal{S}}(s)=[[\alpha]]_{\mathcal{K}}(f(s))$ for all formulas $\alpha$. \\
If $p$ is a propositional variable then by using the definition of fuzzy geometric model morphism, we can show that $[[p]]_{\mathcal{S}}(s)=[[p]]_{\mathcal{K}}(f(s))$ i.e., $gr(s\models p)=gr(f(s)\models p)$.\\
It can be easily shown that $gr(s\models \displaystyle\bigvee_{j\in J}\alpha_j)=gr(f(s)\models\displaystyle\bigvee_{j\in J}\alpha_j)$ and $gr(s\models\alpha_1 \wedge\alpha_2)=gr(f(s)\models \alpha_1\wedge\alpha_2)$. The only part we have to show is that $gr(s\models\heartsuit^{\lambda}(\alpha_1,\alpha_2,\cdots,\alpha_n)=gr(f(s)\models \heartsuit^{\lambda}(\alpha_1,\alpha_2,\cdots,\alpha_n))$. Since $f$ is the coalgebra morphism, henceforth $Tf\circ \sigma_1=\sigma_2\circ f$. So, the diagram shown in Fig. \ref{fig:1} commutes.
\begin{figure}[H]
\begin{center}
\begin{tikzcd}
S \arrow[rr, "f"] \arrow[d, "\sigma_1"'] &  & K \arrow[d, "\sigma_2"] \\
TS \arrow[rr, "Tf"']                     &  & TK                     
\end{tikzcd}
\caption{Coalgebra morphism}
\label{fig:1}
\end{center}
\end{figure}
Now, the following diagram also commutes which is obtained by applying the functor $\mathcal{Q}$ to the above diagram (Fig. \ref{fig:1}).
\begin{figure}[H]
\begin{center}
\begin{tikzcd}
\mathcal{Q}S                                                   &  & \mathcal{Q}K \arrow[ll, "\mathcal{Q}f=f^{-1}"']                                                         \\
\mathcal{Q}(TS) \arrow[u, "\mathcal{Q}\sigma_1=\sigma_1^{-1}"] &  & \mathcal{Q}(TK) \arrow[ll, "\mathcal{Q}(Tf)=(Tf)^{-1}"] \arrow[u, "\mathcal{Q}\sigma_2=\sigma_2^{-1}"']
\end{tikzcd}
\end{center}
\end{figure}
We observe that
\begin{flalign*}
&[[\heartsuit^{\lambda}(\alpha_1,\alpha_2,\cdots,\alpha_n)]]_{\mathcal{S}}(s)& \\ &=\lambda_{S}([[\alpha_1]]_{\mathcal{S}},\cdots,[[\alpha_n]]_{\mathcal{S}})\circ \sigma_1 (s)&\\
                                                                                                                    & = \lambda_{S}([[\alpha_1]]_{\mathcal{K}}\circ f,\cdots,[[\alpha_n]]_{\mathcal{K}}\circ f)\circ \sigma_1 (s) \text{ [ as $f^{-1}([[\alpha]]_{\mathcal{K}})=[[\alpha]]_{\mathcal{S}}$ ] }&\\
                                                                                                                    &= \lambda_{K}([[\alpha_1]]_{\mathcal{K}},\cdots,[[\alpha_n]]_{\mathcal{K}})\circ Tf\circ \sigma_1 (s) \text{ [ by naturality of $\lambda$ ] }&\\
                                                                                                                     &= \lambda_{K}([[\alpha_1]]_{\mathcal{K}},\cdots,[[\alpha_n]]_{\mathcal{K}})\circ \sigma_2 \circ f (s)&\\
                                                                                                                      &= [[\heartsuit^{\lambda}(\alpha_1,\cdots,\alpha_n)]]_{\mathcal{K}}(f(s))&
\end{flalign*}
Therefore, $gr(s\models\heartsuit^{\lambda}(\alpha_1,\alpha_2,\cdots,\alpha_n))=gr(f(s)\models \heartsuit^{\lambda}(\alpha_1,\alpha_2,\cdots,\alpha_n))$.
\end{proof}
\end{prop}
Applying Proposition \ref{p10}, we obtain the following result.
\begin{prop}\label{BVTHM}
	Behaviourally equivalent states are modally equivalent.
	\begin{proof}
		Let $\mathcal{B}=(B,\sigma_1,\mathcal{V}_B)$ and $\mathcal{B}'=(B',\sigma_2,\mathcal{V}_{B'})$ be fuzzy geometric models for $T$. Consider $b\in B$ and $b'\in B'$ are two states. Suppose, the states $b$ and $b'$ are behaviourally equivalent. We shall show that they are modally equivalent. Since $b$ and $b'$ are behaviourally equivalent in $FMOD(T)$, so there exists an object $\mathcal{C}=(C,\gamma,\mathcal{V}_C)$ in $FMOD(T)$ and morphisms $g:\mathcal{B}\to\mathcal{C}$ and $h:\mathcal{B}'\to \mathcal{C}$ in $FMOD(T)$ such that $g(b)=h(b')$. Now, by Proposition \ref{p10}, we have $gr(b\models \alpha)=gr(g(b)\models\alpha)$ and $gr(b'\models\alpha)=gr(h(b')\models\alpha)$, $\forall \alpha\in \textbf{FGML}(\Sigma)$. As $g(b)=h(b')$, hence $gr(b\models\alpha)=gr(b'\models\alpha)$, $\forall \alpha\in \textbf{FGML}(\Sigma)$. Therefore, the states $b$ and $b'$ are modally equivalent.
\end{proof}
		\end{prop}
\section{Final model}
\label{fm}
In this section, we assume that $T$ is an endofunctor on \textbf{SFuzzy-Top}, the category of sober fuzzy topological spaces, and consider a characteristic fuzzy geometric modal signature $\Sigma$ for the endofunctor $T$. We shall create a final model in $FMOD(T)$ for the endofunctor $T$. Let $\mathcal{B}=(B,\gamma,\mathcal{V}_B)$ be a fuzzy geometric model for $T$.
\begin{defn}
\label{p11}
Any two formulas $\alpha$ and $\beta$ are equivalent in $\text{FMOD}(T)$ iff $gr(b\models \alpha)=gr(b\models \beta)$, $\forall b\in B$. Let $\alpha\equiv \beta$ denote the formulas $\alpha$ and $\beta$ are equivalent.
\end{defn}
Let $[\alpha]$ denote the equivalence class of a formula $\alpha\in \textbf{FGML}(\Sigma)$. Let $\mathcal{E}$ be the collection of equivalence classes of formulas in $\textbf{FGML}(\Sigma)$. We define $gr(b\models [\alpha])=gr(b\models \alpha)$, for any $b\in B$.
We shall show that $\mathcal{E}$ is a frame.
\begin{prop}
\label{p13}
$\mathcal{E}$ is a frame.
\begin{proof}
The order relation on $\mathcal{E}$ is defined as: $[\alpha]\leq [\beta]\iff gr(b\models \alpha)\leq gr(b\models \beta)$, $\forall b\in B$. As $gr(b\models \alpha)=gr(b\models\alpha)$, the order relation $\leq$ is reflexive. It is easy to see that if $[\alpha_1]\leq [\beta_1]$ and $[\beta_1]\leq [\beta_3]$ then $[\alpha_1]\leq [\beta_3]$. Thus the order relation $\leq$ is transitive. Now, if $[\alpha]\leq [\beta]$ and $[\beta]\leq [\alpha]$ then by the defined order relation we have $gr(b\models \alpha)=gr(b\models\beta)$, $\forall b\in B$. Hence, $\alpha\equiv \beta$. As a result, $[\alpha]=[\beta]$. So the order relation $\leq$ is antisymmetric. Therefore, $\mathcal{E}$ is a poset with this order relation. As $gr(b\models \alpha\wedge\beta)=[[\alpha\wedge\beta]]_{\mathcal{B}}(b)=[[\alpha]]_{\mathcal{B}}(b)\wedge [[\beta]]_{\mathcal{B}}(b)=gr(b\models\alpha)\wedge gr(b\models\beta)$, hence $[\alpha\wedge\beta]\in \mathcal{E}$. As a result, $[\alpha]\wedge[\beta]\in\mathcal{E}$. Similarly, arbitrary join exists in $\mathcal{E}$. We observe that,  $[\alpha]\wedge\displaystyle\bigvee_{j\in J}[\beta_j]=[\alpha]\wedge[\bigvee_{j\in J}\beta_j]=[\alpha\wedge\bigvee_{j\in J}\beta_j]$. Now, we have $gr(b\models \alpha\wedge\bigvee_{j\in J}\beta_j)=gr(b\models\alpha)\wedge gr(b\models \bigvee_{j\in J}\beta_j)=[[\alpha]]_{\mathcal{B}}(b)\wedge [[\displaystyle\bigvee_{j\in J}\beta_j]]_{\mathcal{B}}(b)=[[\alpha]]_{\mathcal{B}}(b)\wedge Sup_{j\in J}\{[[\beta_j]]_{\mathcal{B}}(b)\}=Sup_{j\in J}\{[[\alpha]]_{\mathcal{B}}(b)\wedge [[\beta_j]]_{\mathcal{B}}(b)\}=[[\displaystyle\bigvee_{j\in J}(\alpha\wedge\beta_j)]](b)=gr(b\models \displaystyle\bigvee_{j\in J}(\alpha\wedge\beta_j))$, $\forall b\in B$. Consequently, $\displaystyle[\alpha\wedge\bigvee_{j\in J}\beta_j]=[\bigvee_{j\in J}(\alpha\wedge\beta_j)]$. Henceforth, $\displaystyle[\alpha]\wedge\bigvee_{j\in J}[\beta_j]=[\bigvee_{j\in J}(\alpha\wedge\beta_j)]=\bigvee_{j\in J}[\alpha\wedge\beta_j]=\bigvee_{j\in J}([\alpha]\wedge [\beta_j])$. Therefore, $\mathcal{E}$ is a frame.
\end{proof}
\end{prop}

\begin{defn}
\label{p14}
Let $\mathcal{F}=\it{PT}(\mathcal{E})$. A map $\tilde{f}: B \to \mathcal{F}$ is defined by $\tilde{f}(b)=h_b$, where $h_b$ is a frame homomorphism from $\mathcal{E}$ to $[0,1]$ defined by $h_b([\alpha])=gr(b\models\alpha)$.
\end{defn}
\begin{note}\label{nt-1}
	\label{pp14}
	The mapping $\tilde{f}: B\to \mathcal{F}$ is fuzzy continuous. Let $[\alpha]\in\mathcal{E}$. Then we show that $\tilde{f}^{-1}(\Psi([\alpha]))=[[\alpha]]_{\mathcal{B}}$ by the following:
	\begin{equation*}
		\tilde{f}^{-1}(\Psi([\alpha]))(b)=\Psi([\alpha])\tilde{f}(b)=\tilde{f}(b)([\alpha])=gr(b\models\alpha)=[[\alpha]]_{\mathcal{B}}(b).
			\end{equation*} 
	Hence, $\forall \alpha\in \textbf{FGML}(\Sigma)$, $\tilde{f}^{-1}(\Psi([\alpha]))$ is a fuzzy open set on $B$. Therefore, $\tilde{f}$ is a fuzzy continuous map.
\end{note}

Let $\mathcal{G}=\mathcal{P}\circ T\circ \it{PT}$. Then $\mathcal{G}:\textbf{FRM}\to\textbf{FRM}$ is a functor. Since the category \textbf{S-FRM} of spatial frames is equivalent to the opposite category of \textbf{SFuzzy-Top}, the endofunctor defined on the category $\textbf{S-FRM}$ is a restriction of $\mathcal{G}$. As $\Sigma$ is characteristic, so the collection $\{\lambda_{B}(\widehat{[\alpha_1]},\cdots,\widehat{[\alpha_n]}):\lambda\in\Sigma, \alpha_i\in\textbf{FGML}(\Sigma), \widehat{[\alpha_i]}\in\mathcal{Q}(PT\mathcal{E}), i=1,\cdots, n\}$ generates the frame $\mathcal{G}(\mathcal{E})$. So, an assignment can be defined on the generators of $\mathcal{G}(\mathcal{E})$, and by Remark \ref{spr}, it can be extended to a frame homomorphism from $\mathcal{G}(\mathcal{E})$ to $\mathcal{E}$.
\begin{defn}
\label{p15}
Define a morphism $\xi:\mathcal{G}(\mathcal{E})\to \mathcal{E}$ in $\mathcal{FRM}$ by $\xi(\lambda_{\mathcal{F}}(\widehat{[\alpha_1]},\widehat{[\alpha_2]},\cdots,\widehat{[\alpha_n]}))=[\heartsuit^{\lambda}(\alpha_1,\alpha_2,\cdots,\alpha_n)]$.
\end{defn}
The well-definedness of the morphism $\xi$ is shown by Lemma \ref{wf}.
\begin{lem}\label{wf}
	Suppose $\displaystyle\bigvee_{i\in \Lambda}(\bigwedge_{j\in K_i}\lambda_{\mathcal{F}}^{i,j}(\widehat{[\alpha_1]}^{i,j},\widehat{[\alpha_2]}^{i,j},\cdots,\widehat{[\alpha_{n_{i,j}}]}^{i,j}) )=\bigvee_{r\in \mathcal{I}}(\bigwedge_{s\in \mathcal{J}_r}\lambda_{\mathcal{F}}^{r,s}(\widehat{[\alpha_1']}^{r,s},\widehat{[\alpha_2']}^{r,s},\cdots,\widehat{[\alpha_{n_{r,s}}']}^{r,s}))$, where $\Lambda$ and $\mathcal{I}$ are the arbitrary index sets, and $K_i$, $\mathcal{J}_r$ are the finite index sets. Then, formulas $\displaystyle\bigvee_{i \in\Lambda}(\bigwedge_{j\in K_i}\heartsuit^{\lambda^{i,j}}(\alpha_1^{i,j},\alpha_2^{i,j},\cdots,\alpha_{n_{i,j}}^{i,j}))$ and $\displaystyle\bigvee_{r\in \mathcal{I}}(\bigwedge_{s\in \mathcal{J}_r}\heartsuit^{\lambda^{r,s}}(\alpha_1'^{r,s},\alpha_2'^{r,s},\cdots,\alpha_{n_{r,s}}'^{r,s}))$ are equivalent in $FMOD(T)$.
	\begin{proof}
		We shall show that, for an object $\mathcal{B}=(B,\gamma,\mathcal{V}_B)$ in $FMOD(T)$, $gr(b\models \displaystyle\bigvee_{i \in\Lambda}(\bigwedge_{j\in K_i}\heartsuit^{\lambda^{i,j}}(\alpha_1^{i,j},\alpha_2^{i,j},\cdots,\alpha_{n_{i,j}}^{i,j})))=gr(b\models\displaystyle\bigvee_{r\in \mathcal{I}}(\bigwedge_{s\in \mathcal{J}_r}\heartsuit^{\lambda^{r,s}}(\alpha_1'^{r,s},\alpha_2'^{r,s},\cdots,\alpha_{n_{r,s}}'^{r,s})))$, $\forall b\in B$.\\ Now we observe that,
		\begin{flalign*}
			&\displaystyle\bigvee_{i\in \Lambda}(\bigwedge_{j\in K_i}\lambda_{B}^{i,j}([[\alpha_1^{i,j}]]_{\mathcal{B}},[[\alpha_2^{i,j}]]_{\mathcal{B}},\cdots,[[\alpha_{n_{i,j}}^{i,j}]]_{\mathcal{B}}) )&\\
		&=\displaystyle\bigvee_{i\in \Lambda}(\bigwedge_{j\in K_i}\lambda_{B}^{i,j}(\tilde{f}^{-1}(\Psi([\alpha_1^{i,j}])),\tilde{f}^{-1}(\Psi([\alpha_2^{i,j}])),\cdots,\tilde{f}^{-1}(\Psi([\alpha_{n_{i,j}}^{i,j}])))) \text{ [By Note \ref{nt-1}]}&\\
		&=\displaystyle\bigvee_{i\in \Lambda}(\bigwedge_{j\in K_i}(T\tilde{f})^{-1}(\lambda_{\mathcal{F}}^{i,j} (  \Psi([\alpha_1^{i,j}]),\Psi([\alpha_2^{i,j}]),\cdots, \Psi([\alpha_{n_{i,j}}^{i,j}]))) \text{ [ By naturality of $\lambda$ ]}&\\
		&=(T\tilde{f})^{-1} ( \displaystyle\bigvee_{i\in \Lambda}(\bigwedge_{j\in K_i}\lambda_{\mathcal{F}}^{i,j} (  \Psi([\alpha_1^{i,j}]),\Psi([\alpha_2^{i,j}]),\cdots, \Psi([\alpha_{n_{i,j}}^{i,j}])))) \text{ [ By  Definition \ref{IVP} ]}&\\
		&=(T\tilde{f})^{-1} ( \bigvee_{r\in \mathcal{I}}(\bigwedge_{s\in \mathcal{J}_r}\lambda_{\mathcal{F}}^{r,s} (  \Psi([\alpha_1'^{r,s}]),\Psi([\alpha_2'^{r,s}]),\cdots, \Psi([\alpha_{n_{r,s}}'^{r,s}])) )) \text{ [ By the given hypothesis]}&\\
		&=\bigvee_{r\in \mathcal{I}}(\bigwedge_{s\in \mathcal{J}_r}(T\tilde{f})^{-1}(\lambda_{\mathcal{F}}^{r,s} (  \Psi([\alpha_1'^{r,s}]),\Psi([\alpha_2'^{r,s}]),\cdots, \Psi([\alpha_{n_{r,s}}'^{r,s}])))) \text{ [ By Definition \ref{IVP}]}&\\
		&=\bigvee_{r\in \mathcal{I}}(\bigwedge_{s\in \mathcal{J}_r}\lambda_{B}^{r,s} (  \tilde{f}^{-1}(\Psi([\alpha_1'^{r,s}])),\tilde{f}^{-1}(\Psi([\alpha_2'^{r,s}])),\cdots, \tilde{f}^{-1}(\Psi([\alpha_{n_{r,s}}'^{r,s}])))) \text{ [ By naturality of $\lambda$ ]}&\\
		&=\bigvee_{r\in \mathcal{I}}(\bigwedge_{s\in \mathcal{J}_r}\lambda_{B}^{r,s}( [[\alpha_1'^{r,s}]]_{\mathcal{B}},[[\alpha_2'^{r,s}]]_{\mathcal{B}},\cdots, [[\alpha_{n_{r,s}}'^{r,s}]]_{\mathcal{B}}))&                    
									\end{flalign*}
		Therefore, for a fuzzy geometric model $\mathcal{B}$, we have $gr(b\models \displaystyle\bigvee_{i \in\Lambda}(\bigwedge_{j\in K_i}\heartsuit^{\lambda^{i,j}}(\alpha_1^{i,j},\alpha_2^{i,j},\cdots,\alpha_{n_{i,j}}^{i,j})))=gr(b\models\displaystyle\bigvee_{r\in \mathcal{I}}(\bigwedge_{s\in \mathcal{J}_r}\heartsuit^{\lambda^{r,s}}(\alpha_1'^{r,s},\alpha_2'^{r,s},\cdots,\alpha_{n_{r,s}}'^{r,s})))$, $\forall b\in B$.					
			\end{proof}
	\end{lem}
 So, $(\mathcal{E},\xi)$ is a $\mathcal{G}$-algebra. Now we construct a $T$-coalgebra structure on $\mathcal{F}=\textit{PT}\mathcal{E}$.
\begin{defn}
\label{16}
Consider a morphism $\phi=\eta^{-1}_{T\mathcal{F}}\circ\it{PT}(\xi):\mathcal{F}\to T\mathcal{F}$, where the morphism $\it{PT}(\xi):\it{PT}(\mathcal{E})\to \it{PT}(\mathcal{G}(\mathcal{E}))$ is defined by $\it{PT}(\xi)(h)(\lambda_{\mathcal{F}}(\widehat{[\alpha_1]},\widehat{[\alpha_2]},\cdots,\widehat{[\alpha_n]}))=gr( [\heartsuit^{\lambda}(\alpha_1,\alpha_2,\cdots,\alpha_n)]\in h)$, where $h\in \textit{PT}(\mathcal{E})=\textit{PT}\mathcal{E}$, and the morphism $\eta_{T\mathcal{F}}:T\mathcal{F}\to  \it{PT}(\mathcal{G}(\mathcal{E}))$ is defined by $\eta_{T\mathcal{F}}(h^*)(\lambda_{\mathcal{F}}(\widehat{[\alpha_1]},\widehat{[\alpha_2]},\cdots,\widehat{[\alpha_n]}))=\lambda_{\mathcal{F}}(\widehat{[\alpha_1]},\widehat{[\alpha_2]},\cdots,\widehat{[\alpha_n]})(h^*)$,
where $h^*\in T\mathcal{F}$.
\end{defn}
\begin{note}
\label{17}
Since $T\mathcal{F}$ is a sober fuzzy topological space, so by Theorem \ref{s0}, $\eta_{T\mathcal{F}}$  is an isomorphism. Consequently, the morphism $\phi$ is well-defined.
\end{note}
\begin{defn}
\label{18}
The triple $(\mathcal{F},\phi,\mathcal{V}_{\mathcal{F}})$ is an object in $FMOD(T)$, where $(\mathcal{F},\phi)$ is a $T$-coalgebra and the valuation function $\mathcal{V}_{\mathcal{F}}:\Phi\to \mathcal{Q}(\mathcal{F})$ is defined by $\mathcal{V}_{\mathcal{F}}(p)(\tilde{g})=gr(p\in \tilde{g})=\tilde{g}(p)$, where $p\in\Phi$ and $\tilde{g}\in \mathcal{Q}(\mathcal{F})$.
\end{defn}
\begin{prop}
\label{19}
The mapping $\tilde{f}:\mathcal{B}\to (\mathcal{F},\phi,\mathcal{V}_{\mathcal{F}})$ is a morphism in $FMOD(T)$.
\begin{proof}
We are to show that $\tilde{f}$ is a coalgebra morphism from $\mathcal{B}$ to $\mathcal{F}$, and $\tilde{f}^{-1}\circ\mathcal{V}_{\mathcal{F}}=\mathcal{V}_B$. It is observed that for every propositional variable $p$,
\begin{equation*}
\begin{split}
 \tilde{f}^{-1}\circ\mathcal{V}_{\mathcal{F}}(p)(b) &=\tilde{f}^{-1}(\mathcal{V}_{\mathcal{F}}(p))(b)\\
                                                                  &=\mathcal{V}_{\mathcal{F}}(p)(\tilde{f}(b))\\
                                                                   &=gr(p\in \tilde{f}(b))\\
                                                                   &=\tilde{f}(b)(p)\\
                                                                   &=gr(b\models p)  \text{ [ By Definition \ref{p14} ] }\\
                                                                   &=\mathcal{V}_B(p)(b).
\end{split}
\end{equation*}
Henceforth, $\tilde{f}^{-1}\circ\mathcal{V}_{\mathcal{F}}=\mathcal{V}_B$. To prove $\tilde{f}$ is a $T$-coalgebra morphism, we show that $T\tilde{f}\circ\gamma=\phi\circ\tilde{f}$ i.e., the diagram shown in Fig.\ref{fig:p4} commutes.
\begin{figure}[H]
	\begin{center}
		
\begin{tikzcd}
	B \arrow[rr, "\tilde{f}"] \arrow[d, "\gamma"'] &  & \mathcal{F} \arrow[d, "\phi"] \\
	TB \arrow[rr, "T\tilde{f}"']                   &  & T\mathcal{F}                 
\end{tikzcd}
\caption{Illustration of $T$-coalgebra morphism}
\label{fig:p4}
\end{center}
\end{figure}
Now we observe that,
\begin{flalign*}
	&gr(T\tilde{f}\circ\gamma(b)\in\lambda_{\mathcal{F}}(\Psi([\alpha_1]),\Psi([\alpha_2]),\cdots,\Psi([\alpha_n])))&\\
	 & = gr(\gamma(b)\in (T\tilde{f})^{-1}\circ\lambda_{\mathcal{F}}(\Psi([\alpha_1]),\Psi([\alpha_2]),\cdots,\Psi([\alpha_n])))&\\
	 &=gr(\gamma(b)\in\lambda_B([[\alpha_1]]_{\mathcal{B}},[[\alpha_2]]_{\mathcal{B}},\cdots,[[\alpha_n]]_{\mathcal{B}})) \text{ [Since $\lambda$ is the natural transformation ]}&\\
	&=\lambda_B([[\alpha_1]]_{\mathcal{B}},[[\alpha_2]]_{\mathcal{B}},\cdots,[[\alpha_n]]_{\mathcal{B}})\circ\gamma(b)&\\
	&=[[\heartsuit^{\lambda}(\alpha_1,\alpha_2,\cdots,\alpha_n)]]_{\mathcal{B}}(b)&\\
	&=gr(b\models\heartsuit^{\lambda}(\alpha_1,\alpha_2,\cdots,\alpha_n))&\\
	&=\tilde{f}(b)([\heartsuit^{\lambda}([\alpha_1],[\alpha_2],\cdots,[\alpha_n])]) \text{ [By Definition \ref{p14} ] }&\\
	&=gr(\phi\circ\tilde{f}(b)\in\lambda_{\mathcal{F}}(\Psi([\alpha_1]),\Psi([\alpha_2]),\cdots,\Psi([\alpha_n])) \text{ [ By Definition \ref{16} ] } &
\end{flalign*}
As $T\mathcal{F}$ is a sober fuzzy topological space, hence it is a $T_0$-space (Kolmogorov space). Therefore, we have $T\tilde{f}\circ\gamma=\phi\circ\tilde{f}$.
\end{proof}
\end{prop}
\begin{thm}
	\label{20}
	The fuzzy geometric model $\mathfrak{F}=(\mathcal{F},\phi,\mathcal{V}_{\mathcal{F}})$ in $FMOD(T)$ is a final model in $FMOD(T)$.
	\begin{proof}
We prove it by showing that for an object $\mathcal{B}=(B,\gamma,\mathcal{V}_B)$ in $FMOD(T)$, a unique $T$-coalgebra morphism exists from $\mathcal{B}$ to $\mathfrak{F}$. Following the Proposition \ref{19}, a $T$-coalgebra morphism $\tilde{f}:\mathcal{B}\to \mathfrak{F}$ exists. The only part that remains to be proven here is that $\tilde{f}$ is unique. Consider a morphism $f^*:\mathcal{B} \to \mathfrak{F}$ in $FMOD(T)$. By Proposition \ref{p10}, we have $gr(b\models \alpha)=gr(f^*(b)\models[\alpha])$. Now $gr(\tilde{f}(b)\models[\alpha])=\tilde{f}(b)[\alpha]=gr(b\models\alpha)=gr(f^*(b)\models[\alpha])$. Consequently, $\tilde{f}(b)=f^*(b)$. Therefore, $\mathfrak{F}$ is final in $FMOD(T)$.
	\end{proof}
\end{thm}
By Theorem \ref{20}, we derive the following result.
\begin{thm}\label{MB}
Modal equivalence implies behavioural equivalence.
\begin{proof}
Let $\mathcal{B}=(B,\gamma,\mathcal{V}_{B})$ and $\mathcal{B}_1=(B_1,\gamma_1,\mathcal{V}_{B_1})$ be fuzzy geometric models for $T$. Let $b\in B$ and $b_1\in B_1$ be states. If $b$ and $b_1$ are modally equivalent then we have $gr(b\models\alpha)=gr(b_1\models\alpha)$, for all formulas $\alpha$. By Proposition \ref{19}, there exist morphisms $\tilde{f}:\mathcal{B}\to\mathfrak{F}$ and $\tilde{f}_1:\mathcal{B}_1\to\mathfrak{F}$ in $FMOD(T)$. Using Proposition \ref{p10}, we have $gr(\tilde{f}(b)\models[\alpha])=gr(b\models\alpha)=gr(b_1\models\alpha)=gr(\tilde{f}_1(b_1)\models[\alpha] )
$, for all formulas $\alpha$. Therefore, $\tilde{f}(b)=\tilde{f}_1(b_1)$. Hence, $b$ and $b_1$ are behaviourally equivalent.
\end{proof}
\end{thm}
\begin{rem}
The converse of the statement mentioned in Theorem \ref{MB} is true by Proposition \ref{BVTHM}. Thus, modal equivalence and behavioural equivalence coincide.
\end{rem}
 
\section{Bisimulations for Fuzzy geometric models }
\label{bf}
The aim of this section is to develop bisimulations for fuzzy geometric models for an endofunctor $T$, where $T$ is defined on $\textbf{Fuzzy-Top}$.
\begin{defn}[\cite{goguen1967fuzzy}]
	\label{21}
	Consider that $F$ and $F'$ are any two sets, and $R$ is a relation between $F$ and $F'$. Then, for a subset $E$ of $F$, $R[E]=\{d'\in F': \exists  \text{ e} \in E, eRd'\}$ and for a subset $E'$ of $F'$
	$R^{-1}[E']=\{d\in F: \exists  \text{ e}' \in E', dRe'\}$.
\end{defn}
Let $\mu$ be a fuzzy set on $F$. Then a fuzzy set $R[\mu]$ on $F'$ can be defined by $R[\mu](d')=\displaystyle\bigvee_{d\in F}\{\mu(d):dRd'\}$.
For a fuzzy set $\eta$ on $F'$, we define an inverse image of $\eta$ under the relation $R$ by $R^{-1}[\eta](d)=\displaystyle\bigvee_{d'\in F'}\{\eta(d'): dRd'\}$. It is clear that $R^{-1}[\eta]$ is a fuzzy set on $F$.\\
We define the Aczel-Mendler bisimulation between fuzzy geometric models for $T$.
\begin{defn}
	\label{22}
	Let $\mathcal{B}_1=(B_1,\gamma_1,\mathcal{V}_{B_1})$ and $\mathcal{B}_2=(B_2,\gamma_2,\mathcal{V}_{B_2})$ be two fuzzy geometric models for $T$. Then, a relation $\mathcal{R}\subseteq B_1\times B_2$ is said to be an Aczel-Mendler bisimulation between $\mathcal{B}_1$ and $\mathcal{B}_2$ if for each $(b_1,b_2)\in\mathcal{R}$ and $p\in \Phi$, $\mathcal{V}_{B_1}(p)(b_1)=\mathcal{V}_{B_2}(p)(b_2)$, i.e., $gr(b_1\models p)=gr(b_2\models p)$ and there exists a coalgebra morphism $\gamma^*:\mathcal{R}\to T\mathcal{R}$ for which the projection maps $\pi_1:\mathcal{R}\to B_1$ and $\pi_2:\mathcal{R}\to B_2$ are coalgebra morphisms and satisfy the relations $\gamma_1\circ\pi_1=T(\pi_1)\circ\gamma^*$, $\gamma_2\circ\pi_2=T(\pi_2)\circ\gamma^*$, i.e. the diagram shown in Fig. \ref{AMB} commutes:
	\begin{figure}[H]
		\begin{center}
			\begin{tikzcd}
				B_1 \arrow[d, "\gamma_1"'] & \mathcal{R} \arrow[l, "\pi_1"'] \arrow[r, "\pi_2"] \arrow[d, "\gamma^*"] & B_2 \arrow[d, "\gamma_2"] \\
				TB_1                       & T\mathcal{R} \arrow[l, "T\pi_1"] \arrow[r, "T\pi_2"']                    & TB_2                     
			\end{tikzcd}
		\end{center}
	\caption{Illustration of Aczel-Mendler bisimulation between fuzzy geometric models}
	\label{AMB}
	\end{figure}
\end{defn}
We now introduce a notion of $\Sigma$-bisimulation between fuzzy geometric models for $T$, adapting the ``$\Lambda$-bisimulation'' concepts discussed in \cite{bakhtiari2017bisimulation,gorin2013simulations}.
First, we introduce the notion of coherent pairs.
\begin{defn}
	Assume that $\mathcal{R}$ is a relation between $B$ and $B'$. Let $\pi_1:\mathcal{R}\to B$ and $\pi_2:\mathcal{R}\to B'$ be projection maps. Then, a pair $(\tilde{r_1},\tilde{r_2})$, where $\tilde{r_1}$ and $\tilde{r_2}$ are respectively the fuzzy sets on $B$ and $B'$, is called $\mathcal{R}$-coherent if $\mathcal{R}[\tilde{r_1}]\leq \tilde{r_2}$ and $\mathcal{R}^{-1}[\tilde{r_2}]\leq \tilde{r_1}$.
\end{defn}

\begin{defn}
	\label{23}
	Let $\mathcal{B}_1=(B_1,\gamma_1,\mathcal{V}_{B_1})$ and $\mathcal{B}_2=(B_2,\gamma_2,\mathcal{V}_{B_2})$ be two fuzzy geometric models for $T$. A relation $\mathcal{R}\subseteq B_1\times B_2$ is said to be a $\Sigma$-bisimulation between $\mathcal{B}_1$ and $\mathcal{B}_2$ if for all $(b_1,b_2)\in\mathcal{R}$, $p\in\Phi$ and each pair of fuzzy open sets $(\mu_i,\eta_i)\in \mathcal{Q}(B_1)\times\mathcal{Q}(B_2)$ such that $\mathcal{R}[\mu_i]\leq\eta_i$ and $\mathcal{R}^{-1}[\eta_i]\leq \mu_i$, we have :
	\begin{enumerate}[(i)]
	\item $gr(b_1\models p)=gr(b_2\models p)$, \text{ and }
	 \item $gr(\gamma_1(b_1)\in\lambda_{B_1}(\mu_1,\mu_2,\cdots,\mu_n))=gr(\gamma_2(b_2)\in\lambda_{B_2}(\eta_1,\eta_2,\cdots,\eta_n))$.
\end{enumerate}
\end{defn}
Two states $b_1\in B_1$ and $b_2\in B_2$ are said to be $\Sigma$-bisimilar if there exists a $\Sigma$-bisimulation $\mathcal{R}$ such that $(b_1,b_2)\in\mathcal{R}$.
\begin{lem}
	Consider $\mathcal{R}\subseteq B_1\times B_2$ is a relation that is equipped with the fuzzy subspace topology. Let $\pi_1:\mathcal{R}\to B_1$ and $\pi_2:\mathcal{R}\to B_2$ be projection maps. Then, a pair of fuzzy open sets $(\mu,\eta) \in  \mathcal{Q}(B_1)\times\mathcal{Q}(B_2)$ is $\mathcal{R}$-coherent $\iff$  $\pi_1^{-1}(\mu)=\pi_2^{-1}(\eta)$.
	\begin{proof}
		Suppose the pair of fuzzy open sets $(\mu,\eta)$ is $\mathcal{R}$-coherent. We shall show that $\pi_1^{-1}(\mu)=\pi_2^{-1}(\eta)$. First, we show that $\pi_1^{-1}(\mu)$ is a fuzzy subset of $\pi_2^{-1}(\eta)$, i.e. $\pi_1^{-1}(\mu)\leq \pi_2^{-1}(\eta)$. We notice that $\pi_2(\pi_1^{-1}(\mu))$ and $\mathcal{R}[\mu]$ are both fuzzy sets on $B_2$. It is simple to demonstrate that $\pi_2(\pi_1^{-1}(\mu))=\mathcal{R}[\mu]$.\\
				Now, \begin{equation*}
			\begin{split}
		\pi^{-1}(\mu) &\leq \pi_2^{-1}(\pi_2(\pi_1^{-1}(\mu)))\\
		 &= \pi_2^{-1}(\mathcal{R}[\mu]) \text{ [As $\pi_2(\pi_1^{-1}(\mu))=\mathcal{R}[\mu]$ ]}\\
		 &\leq \pi_2^{-1}(\eta) \text{ [ As $\mathcal{R}[\mu]\leq \eta$ ]}
		 \end{split}
	 \end{equation*}
  	Similarly, we can show that $\pi_2^{-1}(\eta)\leq \pi_1^{-1}(\mu)$. It can be easily proved that if $\pi_1^{-1}(\mu)= \pi_2^{-1}(\eta)$ then the pair $(\mu,\eta)$ is $\mathcal{R}$-coherent.
  	\end{proof}
\end{lem}
Now, we show that $\Sigma$-bisimilar states are modally equivalent.

\begin{cor}
	\label{24}
	Assume that $T$ is an endofunctor on $\textbf{Fuzzy-Top}$. Then $\Sigma$-bisimilarity implies modal equivalence.
	\begin{proof}
		Let $\mathcal{R}$ be a $\Sigma$-bisimulation between fuzzy geometric models $\mathcal{B}_1=(B_1,\gamma_1,\mathcal{V}_{B_1})$ and $\mathcal{B}_2=(B_2,\gamma_2,\mathcal{V}_{B_2})$. Let $b_1\in B_1 \text{ and } b_2\in B_2$ be two states. Suppose $b_1\mathcal{R}b_2$. We shall show that $gr(b_1\models\alpha)=gr(b_2\models\alpha)$, $\forall \alpha\in \textbf{FGML}(\Sigma)$. If $p$ is a propositional variable, then it follows from the definition of $\Sigma$-bisimulation that $gr(b_1\models p)=gr(b_2\models p)$. It can be easily shown that $gr(b_1\models \alpha_1\wedge \alpha_2)=gr(b_2\models \alpha_1\wedge \alpha_2 )$ and $gr(b_1\models\displaystyle\bigvee_{j\in J}\alpha_j)=gr(b_2\models\displaystyle\bigvee_{j\in J}\alpha_j)$, $J$ is an index set. Now, $gr(b_1\models\heartsuit^{\lambda}(\alpha_1,\alpha_2,\cdots,\alpha_n))
		=gr(\gamma_1(b_1)\in\lambda_{B_1}([[\alpha_1]]_{\mathcal{B}_1},[[\alpha_2]]_{\mathcal{B}_1},\cdots,[[\alpha_n]]_{\mathcal{B}_1}))$. By induction principle, we can show that, for each $i=1,2,\cdots,n$, $\mathcal{R}[ [[\alpha_i]]_{\mathcal{B}_1}]\leq[[\alpha_i]]_{\mathcal{B}_2}$ and $\mathcal{R}^{-1}[ [[\alpha_i]]_{\mathcal{B}_2}]\leq[[\alpha_i]]_{\mathcal{B}_1}$. As $\mathcal{R}$ is a $\Sigma$-bisimulation, we have  $gr(\gamma_1(b_1)\in\lambda_{B_1}([[\alpha_1]]_{\mathcal{B}_1},[[\alpha_2]]_{\mathcal{B}_1},\cdots,[[\alpha_n]]_{\mathcal{B}_1}))=gr(\gamma_2(b_2)\in\lambda_{B_2}([[\alpha_1]]_{\mathcal{B}_2},[[\alpha_2]]_{\mathcal{B}_2},\cdots,[[\alpha_n]]_{\mathcal{B}_2})$. Consequently, $gr(b_1\models\heartsuit^{\lambda}(\alpha_1,\alpha_2,\cdots,\alpha_n))=gr(b_2\models\heartsuit^{\lambda}(\alpha_1,\alpha_2,\cdots,\alpha_n))$. Therefore, $b_1$ and $b_2$ are modally equivalent.
	\end{proof}
\end{cor}
\begin{cor}
	\label{25}
	Let $T$ be an endofunctor on $\textbf{Fuzzy-Top}$; let $\Sigma$ be a monotone fuzzy geometric modal signature for $T$. Then Aczel-Mendler bisimulation is a $\Sigma$-bisimulation.
	\begin{proof}
Consider $\mathcal{B}_1=(B_1,\gamma_1,\mathcal{V}_{B_1})$ and $\mathcal{B}_2=(B_2,\gamma_2,\mathcal{V}_{B_2})$ are fuzzy geometric models for $T$. Let $\mathcal{R}$ be an Aczel-Mendler bisimulation between $\mathcal{B}_1$ and $\mathcal{B}_2$. Then the diagram shown in Fig.\ref{fig:AMB3} commutes.
\begin{figure}[H]
\begin{center}
\begin{tikzcd}
B_1 \arrow[d, "\gamma_1"'] & \mathcal{R} \arrow[l, "\pi_1"'] \arrow[r, "\pi_2"] \arrow[d, "\gamma^*"] & B_2 \arrow[d, "\gamma_2"] \\
TB_1                       & T\mathcal{R} \arrow[l, "T\pi_1"] \arrow[r, "T\pi_2"']                    & TB_2                     
\end{tikzcd}
\caption{Illustration of Aczel-Mendler bisimulation between fuzzy geometric models}
\label{fig:AMB3}
\end{center}
\end{figure}
We shall show that, $\mathcal{R}$ is a $\Sigma$-bisimulation. Suppose $b_1\mathcal{R}b_2$, where $b_1\in B_1$ and $b_2\in B_2$ are the states. Since $\mathcal{R}$ is an Aczel-Mendler bisimulation, we have $gr(b_1\models p)=gr(b_2\models p)$, for any propositional variable $p\in \Phi$. Assume that for each pair of fuzzy open sets $(\mu_i,\eta_i)\in \mathcal{Q}(B_1)\times\mathcal{Q}(B_2)$, $\mathcal{R}[\mu_i]\leq\eta_i$ and $\mathcal{R}^{-1}[\eta_i]\leq \mu_i$. \\
Now,     \begin{flalign*}
	 	&gr(\gamma_1(b_1)\in\lambda_{B_1}(\mu_1,\mu_2,\cdots,\mu_n))&\\
	 	&=\lambda_{B_1}(\mu_1,\mu_2,\cdots,\mu_n)(\gamma_1(b_1))&\\
	 	&=\lambda_{B_1}(\mu_1,\mu_2,\cdots,\mu_n)(T\pi_1)(\gamma^*(b_1,b_2)) \text{ [ As $\gamma_1\circ \pi_1=T\pi_1\circ \gamma^* ]$ }&\\
		&=gr(\gamma^*(b_1,b_2)\in (T\pi_1)^{-1}(\lambda_{B_1}(\mu_1,\mu_2,\cdots,\mu_n)))& \\
		&= gr(\gamma^*(b_1,b_2)\in\lambda_{\mathcal{R}} (\mu_1\circ\pi_1,\mu_2\circ\pi_1,\cdots,\mu_n\circ\pi_1)) \text{ [As $\lambda$ is a natural transformation]}&\\
		&\leq gr(\gamma^*(b_1,b_2)\in\lambda_{\mathcal{R}}(\pi_2^{-1}(\pi_2(\pi_1^{-1}(\mu_1)))),\cdots, \pi_2^{-1}(\pi_2(\pi_1^{-1}(\mu_n))) \text{ [As $\lambda$ is monotone] }&\\
				&= gr(\gamma^*(b_1,b_2)\in\lambda_{\mathcal{R}}(\pi_2^{-1}(\mathcal{R}[\mu_1]),\pi_2^{-1}(\mathcal{R}[\mu_2]),\cdots,\pi_2^{-1}(\mathcal{R}[\mu_n]))) &\\
		&\leq gr(\gamma^*(b_1,b_2)\in\lambda_{\mathcal{R}}(\pi_2^{-1}(\eta_1),\pi_2^{-1}(\eta_2),\cdots,\pi_2^{-1}(\eta_n)))\text{[As $\lambda$ is monotone ]}&\\
		&=gr(\gamma^*(b_1,b_2)\in\lambda_{\mathcal{R}}(\eta_1\circ\pi_2,\cdots,\eta_n\circ\pi_2))&\\
		&=gr(\gamma^*(b_1,b_2)\in (T\pi_2)^{-1}(\lambda_{B_2}(\eta_1,\eta_2,\cdots,\eta_n))) \text{ [As $\lambda$ is natural] }&\\
		&= gr(\gamma_2(b_2)\in\lambda_{B_2}(\eta_1,\eta_2,\cdots,\eta_n))&
	\end{flalign*}
	Therefore, $gr(\gamma_1(b_1)\in\lambda_{B_1}(\mu_1,\mu_2,\cdots,\mu_n))\leq gr(\gamma_2(b_2)\in\lambda_{B_2}(\eta_1,\eta_2,\cdots,\eta_n))$. Similarly, it can be shown that $gr(\gamma_2(b_2)\in\lambda_{B_2}(\eta_1,\eta_2,\cdots,\eta_n))\leq gr(\gamma_1(b_1)\in\lambda_{B_1}(\mu_1,\mu_2,\cdots,\mu_n))$. Finally, we have $gr(\gamma_1(b_1)\in\lambda_{B_1}(\mu_1,\mu_2,\cdots,\mu_n))=gr(\gamma_2(b_2)\in\lambda_{B_2}(\eta_1,\eta_2,\cdots,\eta_n))$. Hence, the result follows.
	\end{proof} 
\end{cor}
\begin{thm}
\label{26}
Let $\mathcal{B}=(B,\gamma,\mathcal{V}_B)$ be a fuzyy geometric model for $T$. Then $\mathcal{R}'=\bigcup\{\mathcal{R}:\mathcal{R} \text{ is a $\Sigma$-bisimulation from $\mathcal{B}$ to $\mathcal{B}$ } \}$ is a $\Sigma$-bisimulation from $\mathcal{B}$ to itself.
\begin{proof}
	Consider for each pair of fuzzy open sets $(\mu_i,\mu_i')$ in $B$, $\mathcal{R}'[\mu_i]\leq\mu_i'$ and $\mathcal{R'}^{-1}[\mu_i']\leq\mu_i$. Let $(b_1,b_2)\in\mathcal{R}'$. Then there exists $\mathcal{R}\in\mathcal{R}'$ such that $(b_1,b_2)\in\mathcal{R}$. Since $\mathcal{R}$ is $\Sigma$-bisimulation, we have $\mathcal{V}_B(p)(b_1)=\mathcal{V}_B(p)(b_2)$ i.e., $gr(b_1\models p)=gr(b_2\models p)$. We observe that $\mathcal{R}[\mu_i]\leq\mathcal{R}'[\mu_i]\leq\mu'_i$ and $\mathcal{R}^{-1}[\mu'_i]\leq\mathcal{R}'^{-1}[\mu'_i]\leq\mu_i$. Now, we get $gr(\gamma(b_1)\in\lambda_{B}(\mu_1,\mu_2,\cdots,\mu_n))=gr(\gamma(b_2)\in\lambda_{B}(\mu'_1,\mu'_2,\cdots,\mu'_n))$. It follows that $\mathcal{R}'$ is $\Sigma$-bisimulation from $\mathcal{B}$ to itself.
\end{proof}
\end{thm}
 We emphasize that the above result is important for further progress of this work. In particular, the above result makes it possible for new theoretical developments, such as coinductive proof principles \cite{rutten2019method}.

\section{Conclusion}
\label{f}
This paper uses coalgebraic methods to define modal operators in fuzzy geometric logic. In order to study modalities in fuzzy geometric logic, we have introduced the idea of fuzzy-open predicate lifting for an endofunctor $T$ on \textbf{Fuzzy-Top}. 
The structures referred to as the fuzzy geometric models for $T$, provide the semantics for our coalgebraic logics. 
We have shown that a final model exists in the category $FMOD(T)$ of fuzzy geometric models, where $T$ is an endofunctor on $\textbf{SFuzzy-Top}$. \\
Finally, we have studied bisimulations for fuzzy geometric models. In this study, we have developed the Aczel-Mendler bisimulation concept for fuzzy geometric models, and we have looked at how it relates to the proposed $\Sigma$-bisimulation concept for fuzzy geometric models. In addition, we have demonstrated that $\Sigma$-bisimilarity implies modal equivalence. For an endofunctor $T$ on $\textbf{SFuzzy-Top}$, behavioural equivalence coincides with  modal equivalence. As a result, $\Sigma$-bisimilarity implies behavioural equivalence. \\
We conclude the study by mentioning future research directions.\\
In this study, we have not considered whether the behavioural equivalence concept implies the $\Sigma$-bisimilarity concept. Following the topological predicate lifting notion for an endofunctor on the category of Stone spaces defined in \cite{enqvist2014generalized}, we can define a stronger fuzzy-open predicate lifting notion for endofunctors on $\textbf{Fuzzy-Top}$. Then, it is possible to show that behavioural equivalence implies $\Sigma$-bisimilarity. In future work, we will do it.\\
Considering an endofunctor $T$ on the category of compact fuzzy Hausdorff spaces, we can show the bi-implication between modal equivalence and $\Sigma$-bisimilarity. In this case, it will be fruitful to adopt a stronger notion of fuzzy geometric modal signature $\Sigma$. We will make an effort to do it in future work.\\
We have already observed that if $T$ is an endofunctor on \textbf{SFuzzy-Top} then behavioural equivalence coincides with  modal equivalence. In our future work, we will verify whether behavioural equivalence and modal equivalence coincide if $T$ is an endofunctor on the category of compact fuzzy Hausdorff spaces.

\section*{Declarations}
\begin{itemize}
\item \textbf{Conflict of interest}\\
The authors declare that there is no conflict of interest.
\item \textbf{Availability of data and materials}\\
Data sharing not applicable to this article as no datasets were generated or analysed during the current study.
\end{itemize}

\newpage

\end{document}